\documentclass{aa}
\usepackage{graphicx}
\usepackage{natbib}

\def\ea{{\rm et al.}\ }
\def\e{ electrons~}
\def\rg{ radio ~galaxies~}
\def\qs{ quasars~}
\def\apj{{\it ApJ}\ }

\def\aj{{\it AJ}\ }
\def\aar{{\it Ann.Rev.A\&A}\ }
\def\aaa{{\it A\&A}\ }
\def\aas{{\it A\&AS}\ }
\def\mnras{{\it MNRAS}\ }
\def\nat{{\it Nat}\ }
\begin{document}
\title{Far-infrared/millimetre Emission in 3C~sources
\thanks{
based on observations with ISO, an ESA project with   
instruments funded by ESA Member States   
(especially the PI countries: France, Germany, the Netherlands and the   
United Kingdom) with the participation of ISAS and NASA}
}
   \subtitle{Dust in \rg and \qs}
   \author{Paola Andreani
          \inst{1}
          \and
	  Robert A. E. Fosbury\inst{2}
          \and
        Ilse van Bemmel\inst{3} 
          \and
         Wolfram Freudling\inst{2}
         }
   \offprints{P. Andreani}
   \institute{Osservatorio Astronomico Padova,
Vicolo dell'Osservatorio 5, Padova, I-35122, Italy
\thanks{\emph{Present address:} Max-Planck I. f\"ur Extraterrestrische
Physik, Postfach 1312, 85741 Garching, Germany}
\email{andreani@mpe.mpg.de}
         \and
Space Telescope - European Coordinating Facility
European Southern Observatory,\\
Karl-Schwarzschild-Str. 2,   D-85748 Garching bei M\"unchen,Germany\\
\email{rfosbury@eso.org,wfreudli@eso.org}
         \and
Kapteyn Astronomical Institute,
       P.O.Box 800, 9700 AV Groningen, The Netherlands
\email{bemmel@astro.rug.nl}
             }

   \date{Received March, 2001; accepted }

   \abstract{
We present far-infrared and millimetric observations of a sample of
3C~objects. Millimetre data were taken at 1.25\,mm with the IRAM 30m
antenna feeding the MPIfR bolometer array. Mid-infrared (MIR) and
far-infrared (FIR) photometry were carried out with the ISOCAM and
ISOPHOT cameras on the ISO Satellite. Additional FIR IRAS observations
are also included. We present the entire Spectral Energy Distributions
(SEDs) from the UV to radio and discuss the emitting mechanisms. Two
composite spectra, one for the radio galaxies and one for the radio
quasars, are built in the object rest frame. While the SEDs of the two
classes differ, they are indistinguishable in the MIR and FIR range
where they probably arise as thermal emission from a dusty torus and a
larger-scale (cooler) dust distribution in the host galaxy.
   \keywords{Galaxies: photometry, ISM - Quasars: general -
ISM: dust - radio continuum: galaxies, ISM}
    }
 \maketitle
%

\section{Introduction}
While classical papers on colours and luminosities of radio galaxies
have explicitly ignored the presence of dust on the evidence of the
strong Ly-$\alpha$ emission, blue colours and small scatter in the
K-band Hubble diagram \citep{LL,D89,L89}, more recent work adopts a more
cautious approach and suggests that some of the observed properties can
be interpreted as evidence for the presence of dust distributed on
scales of tens of kpc. For instance, the detection of significant linear
polarization in the UV/blue aligned light is identified with scattered
light from a hidden AGN due to externally illuminated dust which acts as
a very efficient reflector of UV light
\citep{dSA89,S90,CIM93,dSA96,CIMdSA,MdSA,Bes99,Ver01}. \hfill\break
Images taken with HST show evidence for dust lanes in a large fraction
of 3CR galaxies with redshift $ 0.1 \leq z < 0.5 $ and significant extinction
 is found in the central few kpc in some \rg.
Dust disc orientation is
close to being perpendicular to the radio source axis and obscuration
from dust may dominate the appearance of intermediate redshift sources,
influence the apparent morphology of the optical galaxy and determine
the alignment effect seen in powerful \rg \citep{BAUM,deK96,McC97,deK00}.
\hfill\break 
Furthermore, comparison of line strengths arising from the same ions and
involving a common ground state with calculated line ratios provide a
sensitive measure for reddening. Attempts were made to measure
$\frac{Ly\alpha}{H\alpha}$ in a number of high-$z$ \rg ~ and, in most of
the cases, a significant amount of reddening (with $A_V > 0.3$) is found.

On purely theoretical grounds, one expects the production of significant
amounts of dust at early epochs when galaxies were undergoing vigorous
star formation. Thus there is reason to believe that much of the UV
luminosity of high redshift radio sources is reprocessed by dust. The
question of whether there is indeed a large amount of dust associated
with radio galaxies is important not only from the point of view of
galaxy formation and evolution, but it can help us understand the
apparent differences between radio galaxies and radio quasars and as a
test for the Unification Scheme for AGN (van Bemmel, Barthel, de Graauw, 2000).

Direct emission from dust can be detected at FIR and sub-\,mm
wavelengths where the thermal re-radiation from dust grains is expected.
However, on the basis of IRAS data alone for \qs and \rg, it is not
straightforward to constrain the emission mechanism with great
confidence and, in particular, the relative contribution of thermal and
non-thermal components at FIR wavelengths can be addressed only in
sources which are strongly lobe-dominated and therefore, supposedly,
free of any observable beamed radiation \citep{H92,H94,H95,HBH,VBE00}.
In these objects the radio axis is further
away from our line of sight, and the beaming of non-thermal emission
therefore less efficient. Thus if dust is present, one might see it
in these objects.

The detection of HI absorption \citep{U91,CAR98} and CO emission has
confirmed the existence of dense concentrations of both atomic and
molecular gas in the nuclei of some \rg and \qs and indicated that rich
supplies of molecular gas may be ubiquitous in powerful radio objects
detected by IRAS \citep{MSK,MAZ,eva99,eva01}. \hfill\break Detection of
sub-\,mm continuum (with JCMT at 850\,$\mu$m and with IRAM 30m at
1.25\,mm) in \rg provides a picture in which sources with typical
redshift $z < 1$ are, on average, not detected, while those at higher
redshifts ($z > 2.4$) present strong sub-\,mm fluxes, suggesting the
presence of spatially distributed dust with approximately $10^8$
M$_\odot$ in these objects \citep{CK,D94,I95,Bes98,I98,PvdW,arc00}.
On the one hand, this could simply reflect a K-correction effect: the
steep ($S \propto \nu ^{3-4}$) slope of the dust-emitted spectrum
producing an apparent constancy or increase of observed flux, at a fixed
observed frequency, with redshift at constant luminosity. Alternatively,
one could envisage an evolutionary effect with larger dust masses and
luminosities at higher redshifts. In the latter case, the large FIR
luminosities could be produced by large star-formation rates in
extremely gas-rich galaxies, but it can be also due to a selection
effect which favours the detection of those high-z objects with greater
radio power \citep{PvdW,Bes98}.

At low-redshifts, the 3CR sample was the subject of extensive
investigation at FIR wavelengths but the question of the presence of
dust in these objects could not be settled on the basis of IRAS data
alone. \hfill\break Heckman \ea (1992,1994) claim that, if long
wavelength FIR emission is due to thermal re-radiation by circumnuclear
dust, \qs and \rg are expected to show similar outputs of such
radiation, because of the optical thinness of the thermal emission. But
they show that 3C~\qs are more powerful FIR (60\,$\mu$m) sources than
3C~\rg  and this result can be used either to disprove the unification
scheme of \qs and \rg proposed by Barthel (1989), or to state that the
FIR emission from \qs cannot be due to thermal reradiation from dust
unless the emitting region is very optically thick, resulting in
some degree of radiated anisotropy \citep{PK,gra}).

This issue was the subject of subsequent investigations. By
computing the relative contributions from a relativistically-beamed core
and isotropic emission at 60\,$\mu$m, a significant beamed component
from the nuclei of lobe-dominated \qs and \rg {\it in addition} to the
isotropic thermal dust component was found \citep{H95,HBH}. The infrared
output of \rg and even some \qs should also be affected by contributions
from star-formation episodes \citep{VBBY}. \hfill\break Deep
integrations performed with the ISOPHOT camera on the ISO Satellite on a
small sample of \rg and \qs confirm the previously reported FIR excess
in \qs \citep{VBE00}, with this excess extending up to restframe
wavelengths of $\sim$ 130 \,$\mu$m.
In their \rg the relative contribution of dust at FIR wavelengths
with respect to beamed emission was estimated
to be 98\% and emitted by cool dust.
Meisenheimer \ea (2001) on the
basis of 13 detections out of 20 ISOPHOT observations on 10 pairs of
3C~\rg and \qs do confirm the thermal
nature of the infrared emission, but do not confirm the
presence of an infrared excess in quasars.

We present in this paper ISO FIR and groundbased millimetric
observations of a sample of 3C~sources (see Sect. 2), containing both \qs
and \rg. We discuss the origins of the FIR-\,mm emission and the
differences found between the two classes.  Data are presented in Sect. 3
and the results are discussed in Sect. 4.


\section{The sample selection}

The sample discussed in the present work was selected from a larger
sample which was proposed (proposal reference MRC-3CR) for observation
with the ISOPHOT (Lemke \ea 1996) and, for a subset of sources, with the
ISOCAM (Cesarsky \ea 1996) cameras on the ISO Satellite. The 3C~sources
were selected solely on the basis of their visibility with the ISO
satellite in regions of low Galactic cirrus emission. No objects with $z
< 0.3$ were included and the sample contained 50 sources classified as
quasars and 85 as galaxies. The main purpose of the ISO proposal was an
extensive study of the FIR emission mechanism in 3C~sources and a
comparative study of \rg and \qs. These data, supplemented with
millimetric observations, were to be used to disentangle radio
synchrotron emission from the thermal dust emission.
The assembled
sample spans a large range of radio properties and cosmic look-back
times to avoid the well known biases in high frequency selected samples
that arise from relativistic beaming effects, while maintaining a
statistically useful mix of radio galaxies and quasars. This strategy
demands complete samples selected at low frequencies without any biases
in spectral index, radio morphology or angular size. The 3CR sample is
well suited for this problem within the redshift range that it spans ($z
< 2$) and is still the only completely optically identified sample of
low-frequency selected objects. This sampling frequency (178 MHz) does
guarantee selection due to unbeamed, optically thin and thus isotropic,
emission from the radio lobes and so avoids introducing any orientation bias.

After launch, the sensitivity of ISOPHOT proved to be inadequate to
complete observations effectively in the time available. In the present
work we discuss the observations taken with the IRAM 30m antenna,
feeding the MPIfR bolometer array at 1.25\,mm, of 27 of the 3C~sources,
selected purely on the basis of telescope visibility, from the original
ISO proposal. We also present the ISOPHOT observations in the wavelength
range 5--200 \,$\mu$m of 15 of them and ISOCAM observations of 10. Even
though the original ISOPHOT observations were not completed, it is still
possible to address some of the purposes of the proposal in a
self-consistent way. 
  
\section{Observations}

\subsection{1.25\,mm data}

The 1.25\,mm data presented here were taken with the MPIfR 19-channel
bolometer \citep{kre98} at the focus of the IRAM 30m antenna (Pico
Veleta, Spain) during March 1996 and March 1997. The filter set combined
with the atmospheric transmission produces an effective wavelength
around 1.25\,mm; the beam size is $11 ^{\prime \prime}$ (FWHM) and the
chop throw was set at $32^{\prime \prime}$ with a chopping frequency of
2 Hz. The average sensitivity for each channel, limited principally by
atmospheric noise and measured before any sky-noise subtraction was 60
mJy/$\sqrt{Hz}$. The effect of the sky noise on flux measurements could,
however, be substantially reduced by exploiting the correlation between
signals from the different channels using the standard three beam
(beam-switching + nodding) technique. The average rms value was 1 mJy
for a typical integration time on-source of 2000 s.

Atmospheric transmission was monitored by making frequent skydips from
which the derived zenith opacities were 0.09--0.3. Calibration was
performed using Uranus as primary calibrator and Mars and Quasars from
the IRAM pointing list as secondary sources. The different measurements
vary by less than 5 \% for both planets. If we include the uncertainty
in the planet temperature, we estimate an average flux calibration
uncertainty of 10 \%. Pointing was checked each hour and the average
accuracy achieved was better than $ 3 ^{\prime \prime}$.

The data were reduced assuming that the target sources are unresolved,
i.e., having an extent at \,mm wavelengths smaller than the size of the
central channel. The other 18 channels (excluding one which suffered a
large electronic loss) were then exploited to derive a low-noise sky
estimate. The weighted
 average value of the sky, computed using these
outer 17 channels, was subtracted from the signal in the central
channel. Note that this procedure eliminates only that part of the sky
fluctuation with correlation length smaller than the chop throw
($32^{\prime \prime}$), i.e., from fluid motions at {\it short}
wavelengths. However, the dominant part of the atmospheric noise is
produced by motions of convective shells on large scales (at {\it long}
wavelengths) (Church, 1995), while high frequency (5--20 Hz)
fluctuations do not contribute significantly to the noise and are
averaged over 0.25 s per phase, while the wobbling of the secondary
smooths out the low frequency noise. According to Andreani et al.
(1990), the correlation length for convective shells at these
wavelengths is of the order of several tens of centimeters, i.e., only
fluctuations generated at an altitude greater than 2000m above the
telescope survive the double-switching subtraction and contribute to the
noise. At altitudes greater than 5000m above sea level, however, the
residual water vapour is very low and contributs little to the noise.

1.25\,mm fluxes with the associated 1$\sigma$ uncertainties are given in
table 1. The first uncertainty is statistical while the second one
derives from the calibration uncertainty related to the planetary
measurements and their temperature uncertainty. Our 1.25\,mm fluxes
towards 3C\,286, 3C\,309.1, 3C\,295 and 3C\,325  differ slightly from
those reported by Meisenheimer \ea (2001). They agree, however, within
the calibration uncertainties.

\subsection{IRAS and radio data}

The IRAS data are taken from co-added survey data provided by IPAC and
based on the SCANPI (Scan Processing and Integration Tool) processor.
This procedure computes the one-dimensional co-addition of all the IRAS
survey data of the source. The sensitivity is comparable to that
achieved by the FSC (Faint Source Catalog) for point sources (see the
IPAC manual for details). The resulting fluxes are listed in table 2. We
have also applied the SUPERSCANPI procedure, which is similar to SCANPI
but allows the inference of an average value of the FIR emission of the
whole class of object. By dividing our sample into two separate classes,
 \rg ~ and \qs ~, we have applied the SUPERSCANPI procedure to estimate
the weighted-average fluxes at 12, 25, 60 and 100 \,$\mu$m separately
for each. These values are listed in table 2. The same procedure, but
applied to a much larger set of objects, was used by Heckman \ea (1992).
Within our larger uncertainties our results for \rg and \qs are in
agreement. For 3C\,268.3 IRAS data were taken from Hoekstra \ea (1997).
Radio data are taken from NED (NASA Extragalactic Database).

\subsection{ISO data}

Fifteen objects of the present sample were observed with the ISOPHOT
camera. For two objects, 3C\,268.4 and 3C\,280, data were collected for
our programme (RFOSBURY MRC-3CR2) on July 3$^{rd}$ 1996, at 60 and
90\,$\mu$m. Data for 3C\,295, 3C\,309.1, 3C\,325 and additional
observations of 3C\,280 were taken from 5 to 170\,$\mu$m, those of
3C\,286, 3C\,287 from 5 to 100\,$\mu$m  by Chini and have been
independently published by Meisenheimer \ea (2001). These data together
with those of 3C\,268.4 (at 90 and 170\,$\mu$m) 3C\,313 (90\,$\mu$m),
3C\,288.1 (25-170\,$\mu$m), 3C\,352 and 3C\,356 (90 and 160\,$\mu$m)
were taken from the archive and reduced by us with the Phot Interactive
Analysis tool (PIA) version 8.0 \citep{PIA}. All data were taken in
chopping mode except those for 3C\,352 and 3C\,356 which were mapped in
raster mode. These latter data were reduced as described in van Bemmel
\ea (2000). Data for 3C\,46, 3C\,268.3, 3C\,295, 3C\,337 and 3C\,343.1
are taken from Fanti \ea(2000) and that paper, in which there is an
extensive discussion, should be consulted for any details concerning the
observing procedure and the data reduction.

Our analysis of the ISO data of the sources 3C\,286, 3C\,287, 3C\,295,
3C\,309.1, 3C\,325 agree within the error bars with that by Meisenheimer
\ea (2001) but clear detections towards 3C\,280 were obained because of
the different observational setup with much longer exposure time.

Our reduction procedure is briefly summarized as follows:
\begin{itemize}
\item
Ramps (V/s) of the Edited Raw Data (ERD) are linearized and cosmic hits
are removed using the two-threshold deglitching procedure. A linear fit
is then applied to the ramps providing the Signal per Ramp Data (SRD).
\item
All the data corresponding to one chopper position are then averaged and
extrapolated in time to a value of constant detector response ({\it
Pattern Analysis}).
\item
To correct the remaining cosmic hits, another deglitching procedure is 
applied. All the ramp signals per raster point are averaged and data are
then processed to Signal per Chopper Plateau (SCP) level.
\item
Data are then corrected for reset interval and dark current is
subtracted. A signal linearization which corrects for a changing
response during the measurements is then applied.
\item
The same procedure is applied to the Fine Calibration Source (FSC) data
to provide the internal calibration data. 
\item
Source signal is estimated by subtracting the background. This latter is
determined by averaging the signals corresponding to off-source
positions of the chopper. 
\end{itemize}

The resulting fluxes are listed in table 3. Note that for 3C\,352 and
3C\,356, although our analysis shows a signal-to-noise ratio larger than
7 at 160\,$\mu$m and larger than 5 at 90\,$\mu$m, respectively, a good
estimate of the flux levels cannot be achieved because of excessively
noisy FSC data, which prevent us from performing a good internal calibration.

\subsection{ISOCAM data}

ISOCAM observations were taken towards 3C\,288.1, 3C\,295, 3C\,305,
3C\,309.1, 3C\,343.0, 3C\,343.1, 3C\,345, 3C\,352 and 3C\,356. These
ISOCAM observations together with other radio galaxies and data
reductions will be published elsewhere (Siebenmorgen \& Freudling, 2001,
in preparation). A complete ISOCAM catalogue was built by these authors.
They have reduced staring and raster observations of 3C\,sources in the
ISOCAM archive using a homogeneous procedure. The procedure was
optimized for faint sources, and particular effort was taken to
effectively remove `glitches' in the data. Aperture Photometry was
carried out on all sources using two independent procedures for
background estimates and weighting. We address the reader to that paper
for all further details.

\subsection{UV, optical, near-IR and sub-\,mm data}

Optical and UV and near IR photometry are taken from Best \ea (1998)
deVries \ea (1998) and and deKoff \ea (1996). Two objects, 3C\,322 and
3C\,356 were observed with SCUBA at 450 and 850\,$\mu$m by Archibald \ea (2001).

   \begin{figure*}
   \centering
   \includegraphics[width=16cm]{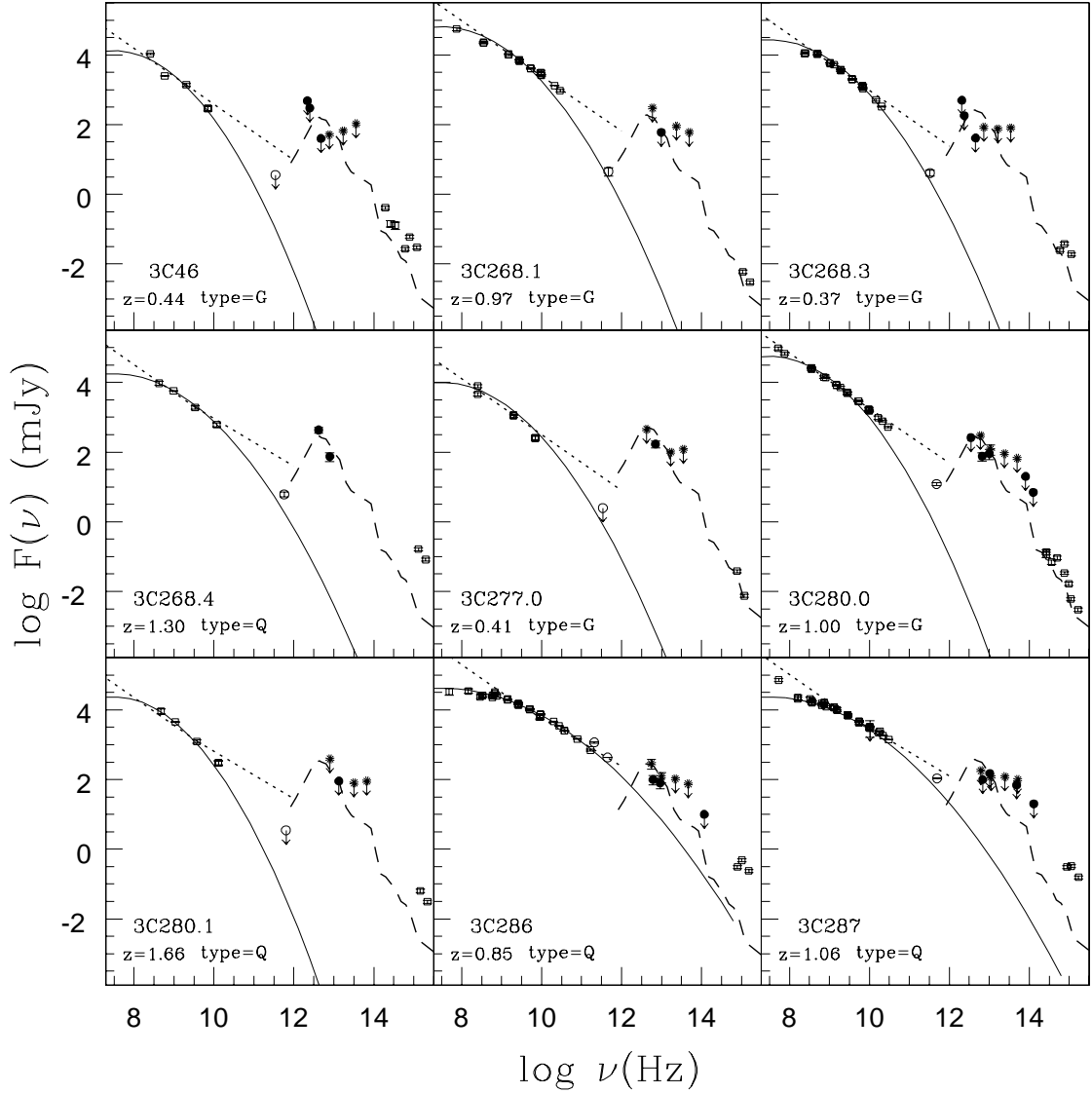}
	\caption{Spectral Energy Distributions (SEDs) from radio to UV
wavelengths. \,mm points are shown as open circles, ISO data as filled
circles, IRAS points at 12, 25, 60 and 100\,$\mu$m are shown as
asterisks, radio and optical observations as open squares. For fluxes
with signal-to-noise ratios, $\frac{S}{N} \leq 3$ a 3$\sigma$ upper
limit is shown as a down-arrow. Two simple approximations of the
non-thermal emission spectra are shown through the radio data: the solid
curve corresponds to equation~\ref{eq:parabola} while short-dashed lines
to the classical power-laws with spectral index of $\alpha_{rg} = -1.04
$ for \rg and $\alpha_{qs} = -0.91$ for quasars. The long-dashed curve
is the output from a model whose basic prescription is the reprocessing
by ISM dust of the UV-optical light (see text for details).}
              \label{FigSED1}%
    \end{figure*}
   \begin{figure*}
   \centering
   \includegraphics[width=16cm]{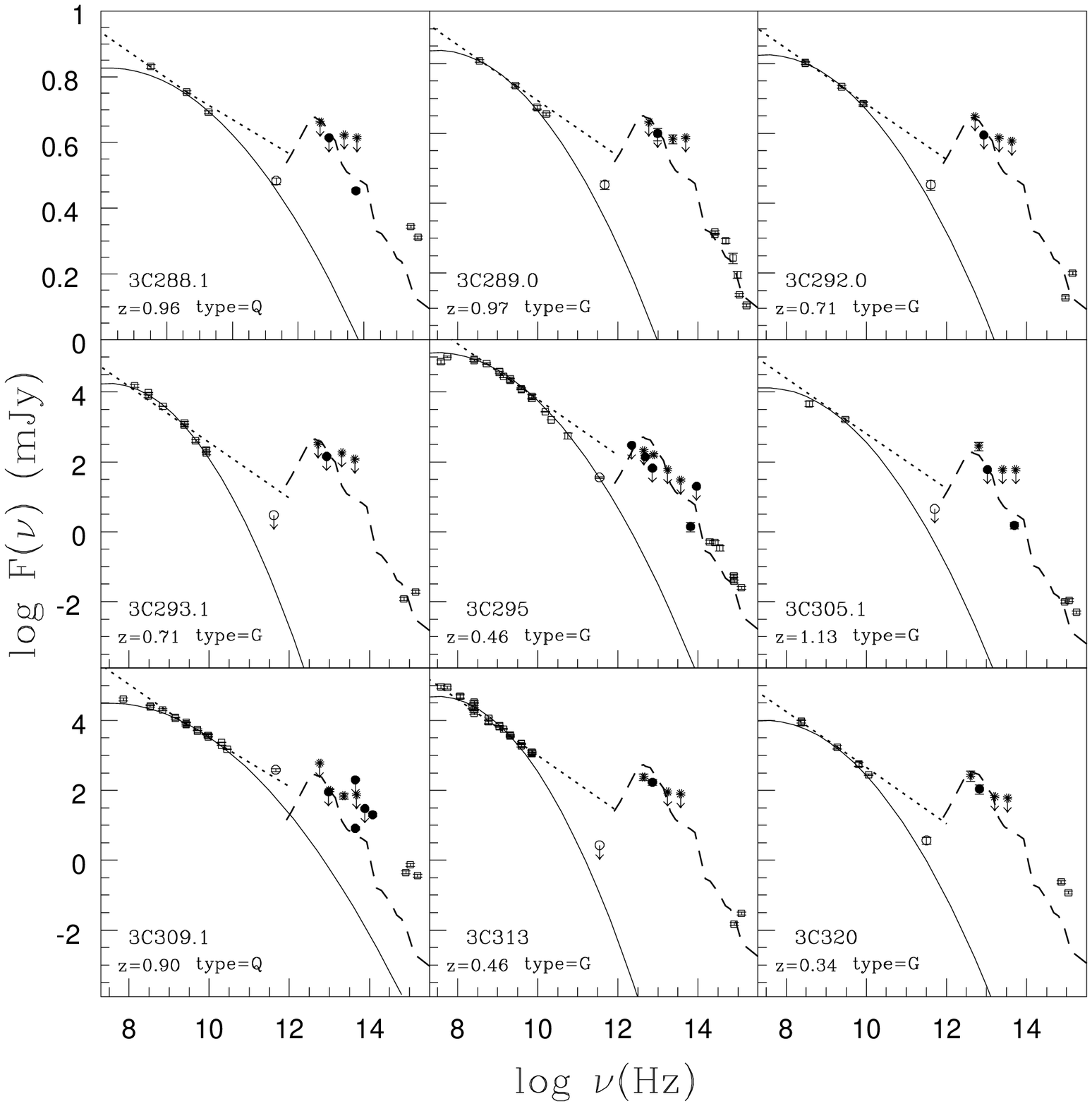}
   \caption{same as figure 1
}
              \label{FigSED2}%
    \end{figure*}
   \begin{figure*}
   \centering
   \caption{same as figure 1
}
   \includegraphics[width=16cm]{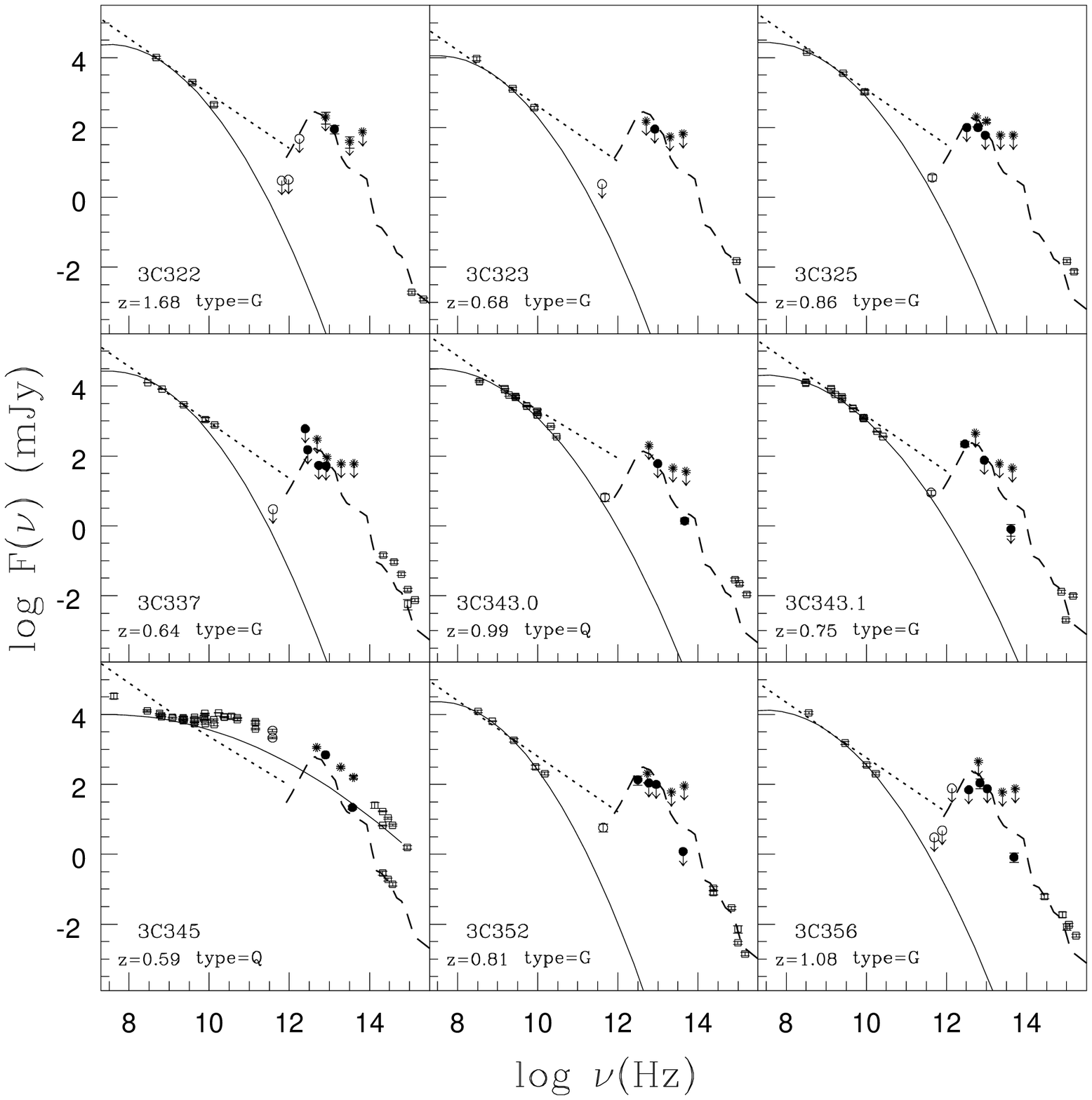}
              \label{FigSED3}%
    \end{figure*}


\section{The Spectral Energy Distribution of Radio Galaxies}

Fig.s 1--3 show the spectral energy distribution from UV to radio
wavelengths of the sample sources. 17 out of 27 objects were detected in
the MIR and/or in the FIR.  We consider as reliable detections all the
MID and FIR fluxes with $\frac{S}{N} \geq 5$, and as marginal detections
those with lower $\frac{S}{N} $ ratio ($3 < \frac{S}{N} <5$).

\subsection{Radio spectrum}
We first consider the radio region of the SED. There is no way to
constrain uniquely the non-thermal radio spectrum unless high-frequency
($\nu > 15$ GHz) radio data are available. A break frequency, depending
on the magnetic field, for a synchrotron spectrum is expected at high
frequencies because of \e cooling. The amount of steepening cannot be
estimated with the present data since most of the sources were not
observed in the frequency range between 5 GHz and 240 GHz (1.25\,mm) and
only one third have data at 15 GHz. It is, therefore, not possible to
reliably extrapolate the radio data at higher frequency to disentangle
any thermal contribution from the 240 GHz flux and infer or reject the
presence of another component contributing to the \,mm emission. For the
17 objects detected in the FIR, it is possible to combine the FIR data
with the \,mm point and try to estimate the two likely contributors ---
thermal and non-thermal.

In what follows, we make two different assumptions. First we assume that
the sources are all sufficiently old that the turn-over frequency is
low. In this case, most of the \e have lost their energy since there is
no mechanism to continuously produce and/or accelerate them. We only fit
points at frequencies above the turn-over (assumed to be at 178 MHz)
with the usual power-law of synchrotron emission
$F \propto \nu ^\alpha$. We take as $\alpha$ the
average values for 3C~\rg and \qs, i.e.  $\alpha _{\rm rg} = 1.04 $
$\alpha _{\rm qso} = 0.91 $ found by \citep{H92} in the frequency range
1.4--15 GHz. We then assume that the synchrotron spectrum between a few
GHz and hundreds of GHz maintains the same slope and we extrapolate it
to \,mm wavelengths. Any difference between the predicted and observed
values at 240 GHz is attributed to an additional spectral component.

This simple approximation provides a poor match to the whole radio
spectrum, low and high frequency radio data lie above and/or below the
expected power-law. In particular, the {\it \,mm emission for all \rg
and the three \qs 3C268.4, 3C\,280.1, 3C\,343.1} lies well below this
extrapolation (see also van Bemmel \& Bertoldi, 2001). It must be
stressed here that the 1.25\,mm observations only refer to the flux
emitted by the source within the central 11$^{\prime \prime}$. If the
radio-lobes dominate the radio spectrum and have a larger extent, it is
possible that some 1.25\,mm flux is lost in our observations. This could
be true for the giant \rg 3C\,277 and for 3C\,356. For all the other
objects, which have a more compact morphology, the further steepening of
the spectrum is real.
\hfill\break
At present the contribution from radio-lobes to the mm flux
it is not observationally settled. For instance,
van Bemmel and Bertoldi (2001) do not
detect any difference in radio-millimetre SED of large and small objects
and suggest that at least in their objects is the core dominating the
millimetre emission.

More generally, a self-absorbed synchrotron emission spectrum
with an electron power-law energy distribution $ \rm N(E) \propto \rm
E^{-s}$, can be parameterised as follows (see e.g. Polletta \ea 2000):

\begin{equation} \label{eq:sync}
L_\nu \propto (\frac{\nu}{\nu_t})^{\alpha_1} ~ 
\{ 1-exp(-(\frac{\nu}{\nu_t})^{\alpha_1-\alpha_2})\}
~e^{-\frac{\nu}{\nu_c}}
\end{equation}

\noindent
where $\alpha_1 $ and $\alpha_2$ are the spectral indices for the
optical thick and thin cases respectively ($\alpha_1 =2.5 $ and
$\alpha_2 = -(s-1)/2$ for a homogeneous source), $\nu_t$ is the frequency
at which the plasma optical depth reaches unity and $\nu_c$ is the
cut-off energy of the plasma energy distribution. For frequencies lower
than $\nu _t$ and by neglecting the high-energy cut-off, $\nu _c$ --
which although depending on the assumed electron energy and the
source magnetic field for relativistic electron and a Galactic magnetic
field it is higher than a few hundreds GHz --
this expression can be approximated by a
parabola-shaped curve in a log-log plane:

\begin{equation} \label{eq:parabola}
\log F_\nu = C + \frac{1}{2 A} (\log \nu - \log
\nu_t)^2\end{equation}

\noindent
where $\frac{1}{A} = 2 \alpha_1 - \alpha _2$. We then use this latter
expression to fit all the radio data with $\nu \leq 240$ GHz where $\nu
_t$ is taken equal for all sources, 31.6 MHz. The free parameters are
$C$, the normalization factor, and $A$ representing the bending of the
curve whose best-fit value varies between -0.2 and -0.55.
For those objects with more than 4 points at frequencies lower than
240 GHz a minimizing-$\chi^2$ fit was carried out, while for those sources
with fewer
than 4 radio points the fit is not statistically significant and
must be considered only as indicative.

These parabola-shaped curves are shown in Fig.s 1--3 as a solid line,
while the power-law curve is shown as short-dashed.

For 3C\,268.3, 286, 293.1, 295, 343, 343.1 a parabola-shaped
non-thermal spectrum fits well the entire radio spectrum up to 240 GHz.
Again we stress here that in most cases and in particular for these
objects the 240 GHz point undoubtedly shows that the radio spectrum
bends at high energy. 
However, it is difficult to disentangle from the \,mm data any residual
emission not related to the synchrotron one but to a thermal cool dust
component.
Even for those objects with well sampled radio spectrum
only the combination of \,mm with FIR data allows the inference of a different emitting
mechanisms on a more solid basis.

3C\,286, 287, 309.1 and 345 have a very large \,mm flux. The overall
spectrum from radio to optical wavelengths could be interpreted as due
to a dominant non-thermal component. As expected, it is not possible to
fit the entire spectrum with a single component, as shown in Fig.s
1--3. It is likely that the \e population does not have a single age for
the entire source. Furthermore, variability plays a role  at least for
3C\,345. Even for 3C\,286 and 3C\,309.1 with FIR
detections it is tough to infer any additional, most
likely thermal, component from cold dust contributing to the
mm flux.

\subsection{FIR and optical spectrum}

For wavelengths $\lambda < 200 \mu m$, the interpretation of the
observed SED is not straightforward since both stars and AGN, each with
their associated dust obscuration and emission, must be taken into
account. To effectively constrain all these components requires high
quality data between 3 and 300 \,$\mu$m in addition to those in the
NIR-optical range. Existing observations allow the construction of
useful SED for thirteen sources with FIR data: 3C\,280 (IRAS/ISO),
3C\,286 (ISO), 3C\,288.1 (ISO), 3C\,289 (IRAS), 3C\,295 (ISO), 3C\,305
(ISO), 3C\,309.1 (ISO/IRAS), 3C\,322 (IRAS), 3C\,325 (ISO), 3C\,343.1
(ISO), 3C\,345 (IRAS/ISO), 3C\,352 (ISO) and 3C\,356 which have reliable
measurements in this range.

Although the number of data available for most of the sources is small,
we use them to compare the observed SED from the optical to 1\,mm with
the model developed by Mazzei and De Zotti (1996) for \rg. They have
constructed a spectrophotometric population synthesis model
incorporating dust extinction and re-emission and a non-thermal central
source which successfully reproduces the SEDs of high-z radio galaxies.

No attempt is made to {\it fit} this model to the \rg data. We do,
however, plot it in Fig.s 1--3, arbitrarily normalized at the
60\,$\mu$m flux (or upper limit). In particular this spectrum was used
to reproduce the radio galaxy SEDs where the {\it scattered} AGN
component dominates the UV rest frame light and the old stellar
population of the host galaxy is the major contribution at longer
wavelengths. The dust reprocessed starburst light
dominates the spectrum at long wavelengths
($\lambda > 10\,\mu$m), while the reprocessing of
AGN energy is neglible. This is consistent with the model in Vernet \ea
(2001) for the z$\sim2.5$ radio galaxy, 4C+48.48.
\hfill\break
Note that the shape of the predicted spectrum agrees quite well with the
observed behavior of the \rg UV-optical-IR SED, in particular for
3C\,280, 3C\,289, 3C\,295, 3C\,305, 3C\,322, 3C\,343.1, 3C\,352 and
3C\,356. Quasars, on the other hand, have a different UV-optical
spectrum, brighter by one dex and dominated by the AGN component.

\subsection{The composite spectra}

To better address the relative importance of the different components in
\rg and \qs, their composite spectra were built and are shown in Fig.
4. The sample was split into two: \rg and quasars. 3C\,345 was not
included in the latter class because of its extreme variability.
In order to compare spectra of sources at different distances data were
normalized to the 60\,$\mu$m flux. The
frequency range is divided into bins in the object rest-frame and the
60\,$\mu$m-normalized data are (weighted) averaged in each bin and errors are
given by the dispersion around the average. When censored
(upper limit) data are present, the Kaplan-Meier estimator was used to
estimate the average values
and their errors \citep{fei,lav92}. Here we would like to
point out the limits of this analysis.
It could be argued that the mid-far-IR spectra of QSOs do not form
a very homogenous classs and on the basis of the present data
alone there seem indeed to be a difference between 268.4, 280.1, 288.1, 343.0,
on the one hand, and 286, 287, 309, 345, on the other hand,
we feel however that it is premature to argue for a qualitative
difference, because of the poor sampling of the spectra of the former
objects, and we continue to consider the QSO class as a whole.

Inspection of Fig. 4 allows us to infer the following:
\begin{itemize}
\item
The two classes of object show a similar radio spectral shape up to a
frequency of around 300 MHz.
\item
For $\nu > $ 300 MHz the \qs~ spectrum is elevated by a factor larger
than 3 up to 240 GHz (1.25\,mm), which is most likely to be due to a
non-thermal, beamed component \citep{VBB}.
\item
The sharp turn-over of the radio spectrum excludes a significant
contribution of the non-thermal component to the FIR fluxes for the 
\rg and half of the \qs studied in this work.
\item
No evident difference is detected in the FIR part of the spectrum
from 100 \,$\mu$m to 1 \,$\mu$m. This result is
satisfactorily in agreement with the fact that the emission
has a thermal origin, very likely linked to the concurrent Starburst
in the galaxy disc. This is consistent with the result of previous
studies \citep{Pol00}.
\item
The optical and UV part of the spectrum differs by one dex
or more between the two classes. \end{itemize}

In the framework of Unified Scheme of AGN this result is not unexpected
since the difference between the two classes is only an
orientation-dependent effect.
An inspection of Fig. 4, together with results from independent
observations, seems to support this picture: (a) the dusty torus emits
between 10 and 50\,$\mu$m and
heated by the AGN, (b) an additional cool component emitting at
100--140\,$\mu$m and representing the reprocessed stellar energy in the
host environment is clearly seen in \rg and it may be outshone by the
stronger non-thermal beamed component in quasars,
(c) the difference at UV-optical wavelengths between the two classes
is due to the obscuration
of the AGN in the radio galaxies by the optically thick torus. The
optical/UV flux which remains visible in these objects can be a
combination of scattered AGN light and a direct view of a young stellar
population \citep{Ver01}.

%
   \begin{figure}
   \centering
   \includegraphics[width=8cm]{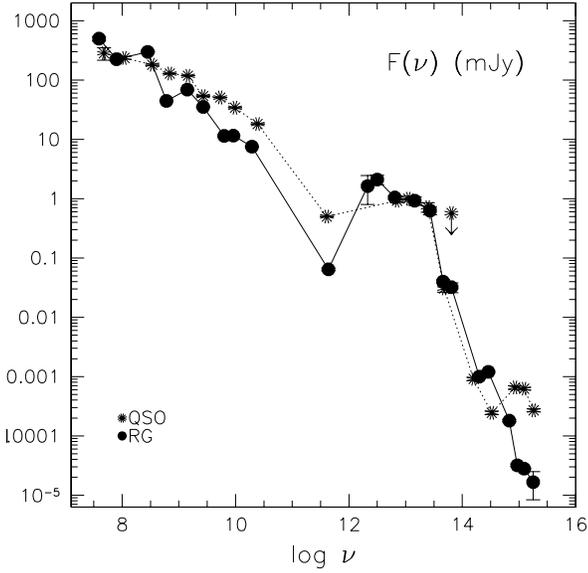}
      \caption{The composite spectra of 3C~\qs and 3C~\rg in our sample.
     \qs points are shown as asterisks, while \rg points as filled
circles. Data are averaged in each wavelength bin in the object's
rest-frame (see text for details). Error bars are in general smaller than
the points size except at $\log \nu$=12.3 (150\,$\mu$m) and at $\log
\nu$=15.25 (0.17\,$\mu$m).
              }
         \label{FigVibStab}
   \end{figure}
%

\section{A colour-magnitude diagram for \rg}

Before analysing flux ratios we show in Fig. 5
1.25\,mm and 60\,$\mu$m fluxes against the source redshift. Both detection
and upper limits are
equally distributed in redshift and do not show any clear trend with $z$.

   \begin{figure}
   \centering
   \includegraphics[width=8cm]{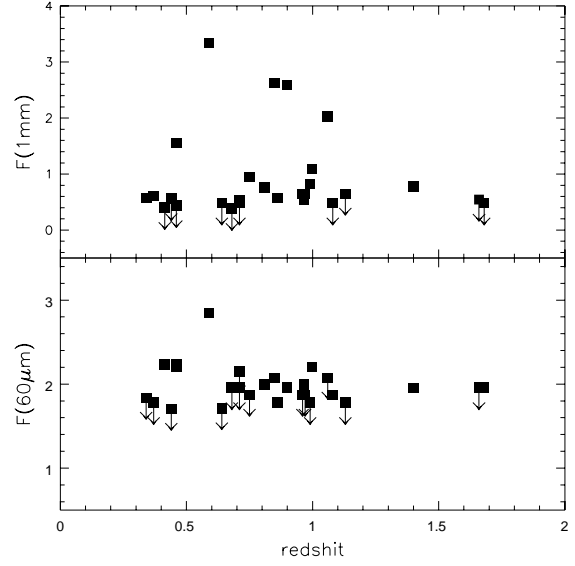}
	\caption{Upper panel reports the 1.25\,mm fluxes, lower panel those at 60\,$\mu$m
against the source redshift. Down arrows correspond to upper limits.
}
         \label{Figzrap}
   \end{figure}

Fig. 6 shows the ratio between the 1.25\,mm and the 60 \,$\mu$m
fluxes, $\frac{F_{1mm}}{F_{60\mu m}}$, against the source redshift. 60
\,$\mu$m was chosen since it is the most common measured wavelength
amongst our objects. Asterisks refer to \qs and filled circles to radio
galaxies. The values for those sources with upper limits at both
wavelengths (3C\,46, 3C\,280.1, 3C\,293.1, 3C\,305.1, 3C\,323, 3C\,337)
are identified as open circles and plotted at the positions of these
limits. Upper limits are given for those objects detected at 60 \,$\mu$m
but not at 1\,mm (3C\,356, 3C\,277, 3C\,313, 3C\,322) and lower limits
for 3C\,268.1, 3C\,268.3, 3C\,287, 3C\,292, 3C\,343, 3C\,343.1,
3C\,288.1 3C\,320, detected at 1\,mm but not in the FIR. Note that there
is no redshift-dependence of the detection rate at either wavelength.
The average value of $\frac{F_{1\,mm}}{F_{60\mu m}}$ for the \rg class
is around 0.05 while for \qs, it is one order of magnitude higher as
seen directly from the relative SEDs (Fig.s 1 through 4). A bimodal
distribution appears in Fig. 6 where objects whose SED is dominated by
the non-thermal emission lie in the upper part of the plot. In those
objects it is likely that any thermal \,mm emission is overwhelmed by
the synchrotron components.

   \begin{figure}
   \centering
   \includegraphics[width=8cm]{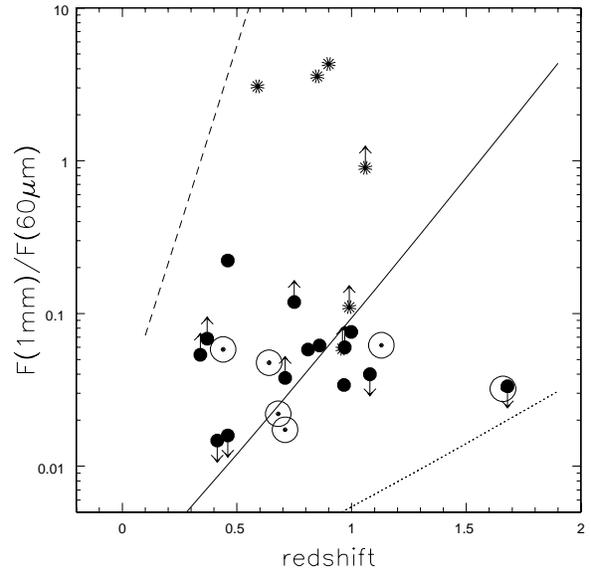}
	\caption{The ratio $\frac{F_{1\,mm}}{F_{60\mu m}}$, is shown against
the source redshift. Filled circles are \rg, asterisks \qs. Open circles
correspond to objects not detected at either wavelength. Variation with $z$
of the ratio for a thermal spectrum, $\epsilon(\lambda) \cdot B(\lambda,T_d)$,
with T$_d$=50 K and $\epsilon \propto \lambda ^{1.5}$ (where $\epsilon$ is
the wavelength-dependent dust emissity)
is represented by a solid line
T$_d$=100 K by the dotted line and T$_d$=20 K by the dashed line.}
         \label{Figzrap}
   \end{figure}

The lower part of the plot contains those sources for which the thermal
component starts to dominate even at \,mm wavelengths. Some of the
objects distribute themselves along a line corresponding to a thermal
spectrum with a dust mean temperature of 50 K at different redshifts.
Objects like 3C\,295, 3C\,268.3, 3C\,320 and 3C\,343.1 fall to the left 
of the plot because of a lower temperature dust component or to a higher
non-thermal component, while 3C\,322, which lies in the lower-right
corner could have a different thermal spectrum with a hotter average
dust temperature (around 100 K). This is indeed supported by the
non-detection of this object in the SCUBA photometry at 450 and
850\,$\mu$m (Archibald et al., 2001).

To further investigate the ratio between the 1.25\,mm and the 60
\,$\mu$m fluxes, $\frac{F_{1\,mm}}{F_{60\mu m}}$, and make it
distance-independent in Fig. 7 $\frac{F_{1\,mm}}{F_{60\mu m}}$ is
plotted against the 60 \,$\mu$m rest-frame luminosity, $L_{60\mu m}$,
for all the sources in our sample. The luminosity is computed in an
Einstein-de Sitter Cosmology with $\Omega=1$ and H$_{\rm 0}=65$km/s/Mpc
and
using a K-corrected thermal spectrum with dust emissivity index of $\beta=1.5$.

   \begin{figure}
   \centering
   \includegraphics[width=8cm]{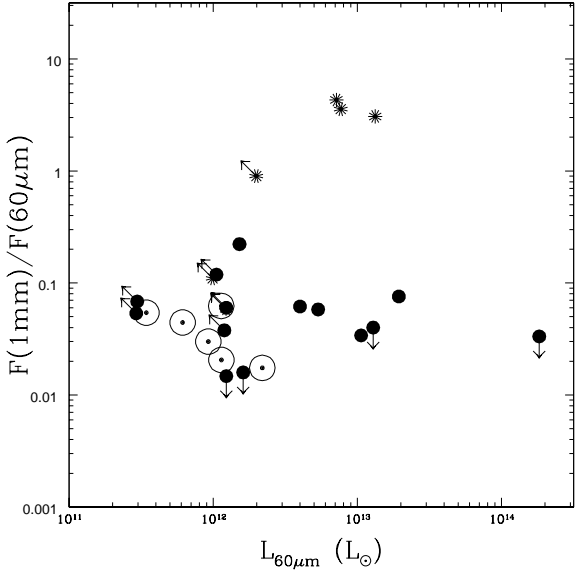}
\caption{The ratio $\frac{F_{1\,mm}}{F_{60\mu m}}$, is plotted against
the source 60 \,$\mu$m luminosity. Filled circles are \rg, asterisks
\qs. Open circles correspond to objects detected at neither wavelength.
}
   \label{Figlumi}
   \end{figure}

While the conclusions drawn from this plot should be regarded as
tentative, the shifting of lower limits, towards the left along the
x-axis and upwards along the y-axis, as indicated by the arrows would
not change the observed trends.
In this figure, the \rg are located preferentially along a sequence of
slightly decreasing $\frac{F_{1\,mm}}{F_{60\mu m}}$ for increasing
$L_{60\mu m}$, while half of the \qs show much larger
$\frac{F_{1\,mm}}{F_{60\mu m}}$ values.

The diagram suggests some interesting conclusions. Half of the \qs share
the common properties of the \rg, while the other half show large
$\frac{F_{1\,mm}}{F_{60\mu m}}$ ratios. This can be explained if these
latter \qs have a dominant non-thermal spectrum, possibly the beamed
component much stronger than that observed for those objects on the
lower-left part of the diagramme. Low value of the ratio
$\frac{F_{1\,mm}}{F_{60\mu m}}$ can be due to the presence of cool dust.
Most of \rg fall in the region where ( $\frac{F_{1\,mm}}{F_{60\mu m}}$,
$L_{60\mu m}$) is consistent with the expected behavior of a thermal
spectrum with temperatures ranging between T=20K and T=50K. Four
objects, 280, 356, 289 and 322, lying to the right in the plot, could be
characterized by a higher dust temperature.
Indeed the ratio
$\frac{F_{1\,mm}}{F_{60\mu m}}$ for a thermal spectrum with
temperature increasing from T=20 K to T=200 K at $z=0.5$ and $z=1.5$
(roughly the redshift range of the present sample) decreases as
the 60\,$\mu$m luminosity increases.

\section{Conclusions}

The investigations of the entire spectral energy distributions of a
small sample of 3C\,sources with \,mm and FIR observations allows us to
draw the following conclusions:
   \begin{enumerate}
\item
Detected fluxes and upper limits at 60\,$\mu$m and 1.25\,mm are
equally distributed in redshift and do not show any trend with $z$.
This result is true also for all the ISO detections.
 
\item
The average power-law spectrum through the radio data, with spectral
indices taken from other studies on \qs and \rg (Heckman et al., 1992),
shows that for most of the objects the
\,mm point lies below the extrapolation of this law to 240 GHz
(1.25\,mm). This means that the radio spectrum
bends at frequencies lower than 240 GHz. In some of the \,mm-detected
objects different spectral components could contribute to the detected
fluxes. The origin of the \,mm emission in these objects is mainly
non-thermal for flat-spectrum sources. In some sources there could be
an additional thermal component from cool dust,
but without sampling the submm region of their spectrum this component still
remains elusive.

\item
QSOs have much stronger \,mm emission. For half of them, the entire
SEDs from radio to UV wavelengths can be due to the superposition of 
different non-thermal components arising from either self-absorbed
synchrotron and/or synchrotron emission from a hard electron spectrum.
However for the remaining \qs, the ratio
$\frac{F_{1\,mm}}{F_{60\mu m}}$ is similar to that of \rg, implying
a common origin of the \,mm-FIR emission.

\item
Composite spectra for \rg and \qs have been constructed in the object
restframe: the main differences are seen in radio - \,mm and optical
SEDs, while the FIR SEDs are remarkably similar. This points to a common
origin of the FIR emission which is likely to be dust-reprocessed energy
from star formation in the host galaxy and/or with a contribution from
the dusty AGN-torus. Although on the basis of these data alone we cannot
fully address this issue in quasars, the far-IR detections in quasars
suggests that there is a thermal as well as a powerful non-thermal
component in these objects.
\item
The similarity of the FIR SED of the two populations is consistent with
the predictions of the orientation-based unified scheme and suggests the
presence of dust either encircling the AGN or in the body of the host
galaxy or both. On the basis of these data alone it is not possible to
identify the main heating mechanism --- AGN or star-formation --- though
the presence of cool dust emission, the width of the FIR spectrum and
modeling suggest that, at least in \rg, dust  --- which could extend to
large distances from the heating source --- is the main FIR emitter.

   \end{enumerate}

\begin{acknowledgements}
P.A. acknowledges support from the Alexander von Humboldt Foundation and
thanks MPE for hospitality. RAEF is affiliated to the Astrophysics
Division of the Space Science Department, European Space Agency.
\hfill\break Part of the data used in this work were taken with the
SCANPI procedure, developed by the NASA Archival center for IRAS
Satellite (IPAC) operating by JPL and made use of the ASURV package Rev
1.2 kindly provided by E. Feigelson. This study has made use of the
NASA/IPAC Extragalactic Database (NED) and was partially supported by
ASI (Italian Space Agency) under contract ARS-98-226 {\it Astrofisica di
sorgenti X e gamma compatte}.
We also thank an anonymous referee for his comments helped in improving this paper.
\end{acknowledgements}

\clearpage

\begin{table*}

\def\nod{...} 
\caption[]{1.25 mm Observations of 3C sources} 

\begin{tabular}{||c|cc|rrr||}
\hline
\noalign{\smallskip}

name & redshift & type & Flux & stat. uncer. & cal. uncer.\\
     &          &      & (mJy)& (mJy) & (mJy)\\

\noalign{\smallskip}
\hline
\noalign{\smallskip}
   3C46   & 0.437 & G  &  $<$3.6 &&\\
  3C268.1 & 0.974 & G  &  4.5    & 1.2 &  0.4\\
  3C268.3 & 0.371 & G  &  4.1    & 0.8 &  0.4\\
  3C268.4 & 1.400 & Q  &  6.1    & 1.0 &  0.6\\
  3C277   & 0.414 & G  &  2.5    & 1.0 &  0.2\\
  3C280.0 & 0.998 & G  & 12.0    & 1.3 &  1.2\\
  3C280.1 & 1.659 & Q  &  3.5    & 1.4 &  0.3\\
  3C286.0 & 0.849 & Q  &428.0    & 3.2 & 40.0\\
  3C287.0 & 1.055 & Q  &108.0    & 1.4 & 10.0\\
  3C288.1 & 0.961 & Q  &  4.4    & 0.9 &  0.4\\
  3C289.0 & 0.967 & G  &  3.4    & 0.9 &  0.3\\
  3C292.0 & 0.713 & G  &  3.4    & 1.1 &  0.3\\
  3C293.1 & 0.709 & G  &  $<$ 3.0&&\\
  3C295.0 & 0.461 & G  & 35.6    & 2.1 &  4.0\\
  3C305.1 & 1.132 & G  &  3.4    & 1.3 &  0.3\\
  3C309.1 & 0.905 & Q  &385.8    & 2.5 & 38.0\\
  3C313.0 & 0.461 & G  &  $<$ 2.7&&\\
  3C320.0 & 0.342 & G  &  3.7    & 0.8 &  0.3\\
  3C322.0 & 1.681 & G  &  $<$ 3.0&&\\
  3C323.0 & 0.679 & G  &  1.9    & 0.8 &  0.2\\
  3C325.0 & 0.860 & G  &  4.3    & 1.0 &  0.4\\
  3C337.0 & 0.635 & G  &  $<$ 3.0&&\\
  3C343.0 & 0.988 & Q  &  6.6    & 1.6 &  0.6\\
  3C343.1 & 0.750 & G  &  8.9    & 1.4 &  0.8\\
  3C345.0 & 0.593 & Q  &3480.0   & 40.0 &300.0\\
          &  & & 2150.0    &  12.0 &100.0\\
  3C352.0 & 0.806 & G  &  5.8    &  1.4 & 0.6\\
  3C356.0 & 1.079 & G  &  $<$ 3.0&&\\
\noalign{\smallskip}
\hline
\hline
\end{tabular}

Uncertainties are given at 1$\sigma$ level\\
upper limits at 3$\sigma$
\end{table*}

\clearpage

\begin{table*}
\def\nod{...} 
             
\caption[]{IRAS Observations of 3C sources} 

\begin{tabular}{||c|cccc||}
\hline
\noalign{\smallskip}

name & 12$\mu m$ & 25 $\mu m$  & 60 $\mu m$ & 100 $\mu m$\\
     & (mJy)     & (mJy)       & (mJy)      & (mJy) \\

\noalign{\smallskip}
\hline
\noalign{\smallskip}
   3C46   & $<$ 105  & $<$  66  & $<$ 141  & $<$ 350\\
  3C268.1 & $<$  60  & $<$  60  & $<$  75  & $<$ 300\\
  3C268.3 & $<$  80  & $<$  74  & $<$ 114  & $<$ 345\\
  3C268.4 &\nod  & \nod    & \nod & \nod \\
  3C277   & $<$ 120  & $<$ 100  & 170$\pm$40& $<$ 450\\
  3C280.0 & $<$  66  & $<$  90  &  90$\pm$30& $<$ 300\\
  3C280.1 & $<$  90  & $<$  75  & $<$  90  & $<$ 390\\
  3C286.0 & $<$  75  & $<$ 105  & $<$  90  & $<$270\\
  3C287.0 & $<$ 100  & $<$ 120  & $<$ 150  & $<$ 180\\
  3C288.1 & $<$  75  & $<$  60  & $<$  75  & $<$ 180\\
  3C289.0 & $<$  75  & 70$\pm$20  &100$\pm$40  & $<$ 210\\
  3C292.0 & $<$  60  & $<$  75  & $<$  90  & $<$ 240\\
  3C293.1 & $<$ 120  & $<$ 180  & $<$ 143  & $<$ 330\\
  3C295.0 & $<$  30  & $<$  60  & $<$  66  & $<$ 210\\
  3C305.1 & $<$  60  & $<$  60  & $<$  60  & 280$\pm$75\\
  3C309.1 & $<$  75  & 80$\pm$14& $<$ 120  & $<$ 600\\
  3C313.0 & $<$  70  & $<$  90  &170$\pm$32&240$\pm$80\\
  3C320.0 & $<$  60  & $<$  66  &110$\pm$33&270$\pm$90\\
  3C322.0 & $<$  60  & $<$  45& 90$\pm$25& 200$\pm$75\\
  3C323.0 & $<$  66  & $<$  54  & $<$  90  & $<$ 150\\
  3C325.0 & $<$  60  & $<$  45  & $<$  60  & $<$ 180\\
  3C337.0 & $<$  60  & $<$  60  & $<$  80  & $<$ 250\\
  3C343.0 & $<$  36  & $<$  45  & $<$  60  & $<$ 180\\
  3C343.1 & $<$  45  & $<$  45  & $<$  75  & $<$ 450\\
  3C345.0 &160$\pm$20&310$\pm$20&700$\pm$30&1140$\pm$70\\
  3C352.0 & $<$  75  & $<$  75  & $<$ 100  & $<$ 240\\
  3C356.0 & $<$  75  & $<$  75  & $<$  75  & $<$ 400\\
          &          &          &          &        \\
\hline
\hline
averages RGs  & 10$\pm$5 & $<$ 15 & 36$\pm$5 & 59$\pm$30\\
averages QSOs & 27.8$\pm$11&23$\pm$8& 29$\pm$8 & 62.6$\pm$27\\
\hline
\end{tabular}

averages were computed with \\
SUPERSCANPI procedures without 3C345\\
\end{table*}

\clearpage

\begin{table*}
\def\nod{...} 
             
\caption[]{ISO Observations of 3C sources} 

\begin{tabular}{||c|cccccccc||}
\hline
\noalign{\smallskip}

name & 5$\mu m$ & 7$\mu m$ & 12$\mu m$ & 25 $\mu m$  & 60 $\mu m$
& 90 $\mu m$ & 170 $\mu m$ & 200 $\mu m$\\
     & (mJy) & (mJy) & (mJy)  & (mJy)  & (mJy) & (mJy) & (mJy) & (mJy) \\

\noalign{\smallskip}
\hline
\noalign{\smallskip}
   3C46   &\nod &\nod &\nod &\nod &$<$50 &$<$40 &$<$300 &$<$480\\
  3C268.3 &\nod &\nod &$<$37 &\nod & $<$84 & $<$41 &$<$180 &$<$500\\
  3C268.4 &\nod &\nod &\nod &\nod &\nod & 74$\pm$20 &430$\pm$ 87 &\nod \\
  3C280.0 &$<$7 &$<$25 &$<$90 & $<$ 120&120$\pm$40 & 76$\pm$21 &$<$260 &\nod \\
  3C286.0 &$<$10 &\nod &\nod &\nod &120$\pm$40 &100$\pm$30 &\nod &\nod \\
  3C287.0 &$<$20 &\nod &$<$70 &\nod &$<$120 &$<$100 &\nod &\nod\\
  3C288.1 &\nod &\nod &2.28$\pm$0.33$^a$ &\nod &$<$180 &$<$136 &$<$190 &\nod\\
  3C295.0 &$<$20 &1.41$\pm$0.43$^a$ &\nod &\nod &160$\pm$40 &140$\pm$35 &\nod &$<$300\\
  3C305.1 &\nod &\nod &1.52$\pm$0.30 $^a$&\nod &\nod &\nod &\nod &\nod\\
  3C309.1 &20$\pm$8&$<$30 &8.18$\pm$0.99$^a$&\nod &100$\pm$30&$<$180 &$<$260 &\nod\\ 
  3C313   &\nod &\nod &\nod &\nod &\nod &$<$200 &\nod &\nod \\
  3C325.0 & $<$50 &$<$20 &$<$60 &\nod &150$\pm$40 &100$\pm$30 &$<$100 &\nod\\
  3C337.0 &\nod &\nod &\nod &\nod &$<$90 &$<$54 &$<$150 & $<$600\\
  3C343.0 &\nod &\nod &1.39$\pm$0.27$^a$ &\nod &\nod & \nod &\nod &\nod\\
  3C343.1 &\nod &\nod &0.80$\pm$0.27$^a$&\nod &\nod &$<$120 &220$\pm$40 &$<$250\\
  3C352.0 &\nod &\nod &$<$1.2 $^a$ &\nod &\nod &$<$110 &135$\pm$70 &\nod\\
  3C356.0 &\nod &\nod &0.83$\pm$0.28 $^a$ &\nod &\nod &110$\pm$65 &$<$70 &\nod\\
          &   & &  & & &     &     &        \\
\hline
\hline
\end{tabular}

$^a$ ISOCAM data
\end{table*}

\end{document}